# Markov-Modulated Linear Regression

Alexander M. Andronov
Dept. of Mathematical Methods and Modelling
Transport and Telecommunication Institute
Riga, Latvia
lora@mailbox.riga.lv

Nadezda Spiridovska
Dept. of Mathematical Methods and Modelling
Transport and Telecommunication Institute
Riga, Latvia
Spiridovska.N@tsi.lv

*Abstract*—Classical linear regression is considered for a case when regression parameters depend on the external random environment. The last is described as a continuous time Markov chain with finite state space. Here the expected sojourn times in various states are additional regressors. Necessary formulas for an estimation of regression parameters have been derived. The numerical example illustrates the results obtained.

*Markov-Modulated processes; linear regression; external environment*

## I. Model description

Classical linear regression [1 – 3] is of the form

$$Y_i = x_i \beta + Z_i, \quad i = 1, \ldots, n, \quad (1)$$

where $Y_i$ is scale response, $x_i = (x_{i,1}, x_{i,2}, \ldots, x_{i,k})$ is $1 \times k$ vector, $\beta$ is $k \times 1$ vector, $Z_i$ is scale disturbance. The usual assumptions take place: the disturbances $Z_i$ are independently, identically normally distributed with mean zero and variance $\sigma^2$, the $n \times k$ matrix $X = (x_{i,v}) = (x_i^T)^T$ has rank $r(X) = k$, so $(X^T X)^{-1}$ exists.

Now we suppose that model (1) corresponds to one unit of a continuous time, $Z_i(t)$ is Brown motion and responses $Y_i(t)$ are time-additive. Then for $t > 0$

$$Y_i(t) = x_i \beta t + Z_i \sqrt{t}, \quad i = 1, \ldots, n, \quad (2)$$

where $Y_i(0) = 0$ and disturbances $Z_i$ are (as before) independently, identically normally distributed with mean zero and variance $\sigma^2$.

Let value $Y_i$ of the $i$-th response be fixed after time $t_i$, so $Y_i = Y_i(t_i)$. In this case the generalized least square method's estimates of $\beta$ and $\sigma^2$ are the following:

$$\tilde{\beta} = (X^T W^2 X)^{-1} X^T Y,$$
$$\tilde{\sigma}^2 = \frac{1}{n-k-1} (Y - W^2 X \tilde{\beta})^T W^{-2} (Y - W^2 X \tilde{\beta}), \quad (3)$$

where $W = diag(\sqrt{t_1}, \ldots, \sqrt{t_n})$.

Additionally we suppose that model (2) operates in the so-called *external environment*, which has final state space $E$ [4]. For the fixed state $s_j \in S$, $j = 1, \ldots, m$, parameters $\beta$ of model (2) are $\beta_j = (\beta_{1,j}, \ldots, \beta_{k,j})^T$, but as before $Z_{i,j}$ are stochastically independent, normally distributed with mean zero and variances $\sigma^2$. Let $t_i = (t_{i,1}, \ldots, t_{i,m})$ be $1 \times m$ vector, for that component $t_{i,j}$ means a sojourn time for response $Y_i$ in the state $s_j \in S$. Note that $t_i = t_{i,1} + \ldots + t_{i,m}$. Then

$$Y_i(t_i) = x_i \sum_{j=1}^{m} \beta_j t_{i,j} + \sum_{j=1}^{m} \sqrt{t_{i,j}} Z_{i,j}, \quad i = 1, \ldots, n.$$

Taking into account properties of the normal distribution, we can rewrite the last formula as

$$Y_i(t_i) = x_i \sum_{j=1}^{m} \beta_j t_{i,j} + Z_i \sqrt{t_i}, \quad i = 1, \ldots, n. \quad (4)$$

To write it in matrix notation we use Kronecker product $\otimes$ [2, 3, 5], the $k \times m$ matrix $\beta = (\beta_1, \ldots, \beta_m) = (\beta_{v,j})$, the $n \times 1$ vector $Z = (Z_i)$, the $1 \times m$ vector $\vec{t}_i = (t_{i,1}, \ldots, t_{i,m})$, the $i$-th rows $e_i$ of $n$-dimensional identity matrix, the $n$-dimensional diagonal matrix $diag(v)$ with the vector $v$ on the main diagonal, *vec* operator *vec A* of matrix *A*. Then

$$Y(T) = (Y_1(t_1), \ldots, Y_n(t_n))^T =$$
$$= \begin{pmatrix} \vec{t}_1 \otimes x_1 \\ \vec{t}_2 \otimes x_2 \\ \ldots \\ \vec{t}_n \otimes x_n \end{pmatrix} vec\beta + diag(\sqrt{t_1}, \sqrt{t_2}, \ldots, \sqrt{t_n}) Z. \quad (5)$$

We see that the generalized linear regression model has place here. The expectation and the covariance matrix of $Y(T)$ are the following:

The article is written with the financial assistance of European Social Fund. Project Nr.2009/0159/1DP/1.1.2.1.2/09/ IPIA/VIAA/006. The Support in Realisation of the Doctoral Programme "Telematics and Logistics" of the Transport and Telecommunication Institute.

$$E(Y(T)) = \begin{pmatrix} \vec{t}_1 \otimes x_1 \\ \vec{t}_2 \otimes x_{21} \\ \ldots \\ \vec{t}_n \otimes x_n \end{pmatrix} vec\beta, \quad (6)$$

$$Cov(Y(T)) = \sigma^2 diag(t_1, t_2, \ldots, t_n).$$

Now we are able to use the generalized least square method to estimate parameter matrix $\beta$, supposing that the matrix of regressors by $vec\,\beta$ has full rank $mk$. If $Y$ means an observed value of $Y(T)$ then

$$vec\widetilde{\beta} = \left( \begin{pmatrix} \vec{t}_1 \otimes x_1 \\ \vec{t}_2 \otimes x_2 \\ \ldots \\ \vec{t}_n \otimes x_n \end{pmatrix}^T \begin{pmatrix} t_1 & 0 & \ldots & 0 \\ 0 & t_2 & \ldots & 0 \\ \ldots & \ldots & \ldots & \ldots \\ 0 & 0 & 0 & t_n \end{pmatrix}^{-1} \begin{pmatrix} \vec{t}_1 \otimes x_1 \\ \vec{t}_2 \otimes x_2 \\ \ldots \\ \vec{t}_n \otimes x_n \end{pmatrix} \right)^{-1}$$

$$\cdot \begin{pmatrix} \vec{t}_1 \otimes x_1 \\ \vec{t}_2 \otimes x_2 \\ \ldots \\ \vec{t}_n \otimes x_n \end{pmatrix}^T \begin{pmatrix} t_1 & 0 & \ldots & 0 \\ 0 & t_2 & \ldots & 0 \\ \ldots & \ldots & \ldots & \ldots \\ 0 & 0 & 0 & t_n \end{pmatrix}^{-1} Y =$$

$$= \left( \sum_{i=1}^n \frac{1}{t_i} (\vec{t}_i^T \vec{t}_i) \otimes (x_i^T x_i) \right)^{-1} \begin{pmatrix} t_1^{-1} \vec{t}_1 \otimes x_1 \\ t_2^{-1} \vec{t}_2 \otimes x_2 \\ \ldots \\ t_n^{-1} \vec{t}_n \otimes x_n \end{pmatrix}^T Y. \quad (7)$$

The variance $\sigma^2$ can be estimated as usually:

$$\widetilde{\sigma}^2 = \frac{1}{n - mk - 1} \left( Y(T) - \begin{pmatrix} \vec{t}_1 \otimes x_1 \\ \vec{t}_2 \otimes x_2 \\ \ldots \\ \vec{t}_n \otimes x_n \end{pmatrix} vec\widetilde{\beta} \right)^T W^{-2} \cdot$$

$$\cdot \left( Y(T) - \begin{pmatrix} \vec{t}_1 \otimes x_1 \\ \vec{t}_2 \otimes x_2 \\ \ldots \\ \vec{t}_n \otimes x_n \end{pmatrix} vec\widetilde{\beta} \right) \quad (8)$$

Note that the last estimate is biased because it doesn't take into account randomness of $\{\vec{t}_i\}$ values.

## II. MARKOV-MODULATED CASE

Further we suppose that the external environment is a random one and is described by a continuous-time Markov chain $J(t), t \geq 0$, with the finite state set $S = \{1, 2, \ldots, m\}$ [4, 6].

Let $\lambda_{i,j}$ be the known transition rate from state $s_i$ to state $s_j$, and $\Lambda_i = \sum_{j \neq i} \lambda_{i,j}$.

We have $n$ independent realizations of this Markov chain. The $i$-th realization corresponds to the response $Y_i$. Let us use the following notation for the $i$-th realization: $N(i)$ be the number of jumps of random environment $J$; $V_1, V_2, \ldots, V_{N(i)}$ be time moments of these jumps and $J_1, J_2, \ldots, J_{N(i)}$ be the corresponding states $J(V_1+), J(V_2+), \ldots, J(V_{N(i)}+)$; $I = J(0)$ be an initial state at time moment 0.

Therefore now sojourn time $T_{i,j}$ in the state $s_j$ for the $i$-th realization is a sum of all terms $V_r - V_{r-1}$, for which $J_r = j$. Above, the case has been considered when the whole trajectory of environment $J(.)$ is known, so $t_{i,j}$ is a fixed value of $T_{i,j}$. We suppose that total time $t_i$ of the $i$-th observation is fixed, so $t_i = T_{i,1} + \ldots + T_{i,m}$. If $\vec{T}_i = (T_{i,1}, \ldots, T_{i,m})$ is the $1 \times m$ vector, then for the general case linear regression (5) is of the form

$$Y(T) = (Y_1(t_1), \ldots, Y_n(t_n))^T =$$
$$= \begin{pmatrix} \vec{T}_1 \otimes x_1 \\ \vec{T}_2 \otimes x_2 \\ \ldots \\ \vec{T}_n \otimes x_n \end{pmatrix} vec\beta + diag(\sqrt{t_1}, \sqrt{t_2} \ldots \sqrt{t_n}) Z \quad (9)$$

Note that $T = \{\vec{T}_1, \ldots, \vec{T}_n\}$ and $Z$ are independent, so the expectation $E(Y(T))$ is the same as before in (6).

Now a special case of the given sample will be considered. It is supposed that for each realization $i$ the following data are available: total observation time $t_i = T_{i,1} + \ldots + T_{i,m}$, initial $J_{i,0}$ and final $J_{i,\tau(i)}$ states of $J(.)$, and the response $Y_i = Y_i(t_i)$ from (9). On this basis we must estimate the unknown parameters: the $k \times m$ matrix $\beta = (\beta_1 \ldots \beta_m) = (\beta_{v,j})$ and the variance $\sigma^2$. Additionally we use a knowledge on parameters of the modulated Markov chain $J(.)$. One allows us to calculate the average sojourn time $E(T_{i,j} | t_i, J_{i,0}, J_{i,\tau(i)})$ in the state $s_j$ during time $t_i$ for the $i$-th realization, given fixed initial and final states $J_{i,0}$ and $J_{i,\tau(i)}$ of $J(.)$, see below Section 3. This time will be used instead of $t_{i,j}$ in previous formulas (7) and (8) so we set $t_{i,j} = E(T_{i,j}|t_i, J_{i,0}, J_{i,\tau(i)})$. Then $E(\vec{T}_i) = E(T_{i,1},...,T_{i,m}) = \vec{t}_i$ and as before

$$vec\widetilde{\beta} = \left(\sum_{i=1}^{n} \frac{1}{t_i}(\vec{t}_i^T \vec{t}_i) \otimes (x_i^T x_i)\right)^{-1} \begin{pmatrix} t_1^{-1}\vec{t}_1 \otimes x_1 \\ t_2^{-1}\vec{t}_2 \otimes x_2 \\ ... \\ t_n^{-1}\vec{t}_n \otimes x_n \end{pmatrix}^T Y. \quad (10)$$

Substitution $Y(T)$ from (9), we get a form which is more convenient for statistical analysis

$$vec\widetilde{\beta} = \left(\sum_{i=1}^{n} \frac{1}{t_i}(\vec{t}_i^T \vec{t}_i) \otimes (x_i^T x_i)\right)^{-1} \begin{pmatrix} t_1^{-1}\vec{t}_1 \otimes x_1 \\ t_2^{-1}\vec{t}_2 \otimes x_2 \\ ... \\ t_n^{-1}\vec{t}_n \otimes x_n \end{pmatrix}^T \cdot$$

$$\cdot \left(\begin{pmatrix} t_1^{-1}\vec{T}_1 \otimes x_1 \\ t_2^{-1}\vec{T}_2 \otimes x_2 \\ ... \\ t_n^{-1}\vec{T}_n \otimes x_n \end{pmatrix} vec\beta + diag(\sqrt{t_1} \; ... \; \sqrt{t_n})Z\right). \quad (11)$$

As $E(\vec{t}_i^T \vec{T}_i | t_i, J_{i,0}, J_{i,\tau(i)}) = \vec{t}_i^T E(\vec{T}_i | t_i, J_{i,0}, J_{i,\tau(i)}) = \vec{t}_i^T \vec{t}_i$, $T$ and $Z$ are independent, and $vecZ$ has zero expectation, we can conclude that the estimate (10) of **$\beta$ is unbiased**.

Further we give necessary formulas for a calculation of the conditional average sojourn time that allows us to get the estimates needed.

### III. MODULATING MARKOV CHAIN

For transition probabilities $p_{i,j}(t) = P\{J(t) = j | J(0) = i\}$ of the above described Markov chain, a usual system of differential equations take place [4, 6]. If $P(t) = (p_{i,j}(t))$ and $\lambda = (\lambda_{i,j})$ are the $m \times m$ matrices, $\Lambda$ is an $m$-dimensional diagonal matrix with a vector $(\Lambda_1,...,\Lambda_m)$ on the main diagonal then

$$\dot{P}(t) = -P(t)\Lambda + P(t)\lambda, \; t \geq 0.$$

The solution can be represented by the matrix exponent [7, 8]:

$$P(t) = \exp(t(\lambda - \Lambda)), \; t \geq 0, \quad (12)$$

where $P(0) = I$.

If all the eigenvalues of matrix $A = \lambda - \Lambda$ are different then the solution (12) can be represented more simply. Let $v_\eta$ and $Z_\eta$, $\eta = 1, ..., m$, be the eigenvalue and the corresponding eigenvector of $A$, $Z = (Z_1,...,Z_m)$ be the matrix of the eigenvectors and $\overline{Z} = Z^{-1} = (\overline{Z}_1^T,...,\overline{Z}_k^T)^T$ be the corresponding inverse matrix (here $\overline{Z}_\eta$ is the $\eta$-th row of $\overline{Z}$). Then [7, 8]:

$$P(t) = \exp(tA) = Zdiag(\exp(v_1 t),...,\exp(v_m t))Z^{-1} =$$
$$= \sum_{\eta=1}^{m} Z_\eta \exp(\gamma_\eta t)\overline{Z}_\eta. \quad (13)$$

For the conditional average sojourn time $t_{r,v}(\tau) = E(T_{r,v}|t_r = \tau, J_{r,0} = i, J_{r,\tau} = j)$ in the state $v \in S$ on the interval $(0, \tau)$ we have

$$t_{r,v}(\tau) = \frac{1}{p_{i,j}(\tau)} \int_0^\tau p_{i,v}(u) p_{v,j}(\tau - u) du. \quad (14)$$

Further

$$p_{i,j}(\tau) = \sum_{\eta=1}^{m} Z_{i,\eta} \exp(\gamma_\eta \tau) \overline{Z}_{\eta,j}, \quad (15)$$

$$\int_0^\tau p_{i,v}(u) p_{v,j}(t-u) du =$$

$$= \int_0^\tau \sum_{\eta=1}^{m} Z_{i,\eta} \exp(\gamma_\eta u) \overline{Z}_{\eta,v} \sum_{\theta=1}^{m} Z_{v,\theta} \exp(\gamma_\theta (\tau - u)) \overline{Z}_{\theta,j} du =$$

$$= \sum_{\eta=1}^{m} Z_{i,\eta} \overline{Z}_{\eta,v} \sum_{\theta=1,\theta\neq\eta}^{m} Z_{v,\theta} \overline{Z}_{\theta,j} \exp(\gamma_\theta \tau) \frac{1}{\gamma_\theta - \gamma_\eta} \cdot$$

$$\cdot (1 - \exp(-\tau(\gamma_\theta - \gamma_\eta))) +$$

$$+ \tau \sum_{\eta=1}^{m} Z_{i,\eta} \overline{Z}_{\eta,v} Z_{v,\eta} \overline{Z}_{\eta,j} \exp(\gamma_\eta \tau) =$$

$$= \tau \sum_{\eta=1}^{m} Z_{i,\eta} \overline{Z}_{\eta,v} Z_{v,\eta} \overline{Z}_{\eta,j} \exp(\gamma_\eta \tau) +$$

$$+ \sum_{\eta=1}^{m} Z_{i,\eta} \overline{Z}_{\eta,v} \sum_{\theta=1,\theta\neq\eta}^{m} Z_{v,\theta} \overline{Z}_{\theta,j} \frac{1}{\gamma_\theta - \gamma_\eta} (\exp(\gamma_\theta \tau) - \exp(\gamma_\eta \tau)).$$

(16)

Now we can make calculation by formula (14) and derive estimates (10) setting $t_{i,v} = t_{i,v}(t_i)$.

### IV. SIMULATION STUDY

Our example supposes three states of the environment ($m = 3$). The known transition rates $\{\lambda_{i,j}\}$ from state $s_i$ to state $s_j$ are set by matrix

$$\lambda = (\lambda_{i,j}) = \begin{pmatrix} 0 & 0.2 & 0.3 \\ 0.1 & 0 & 0.2 \\ 0.4 & 0 & 0 \end{pmatrix}.$$

Stationary state distribution is the following [6]: $\pi = (0.364 \ 0.242 \ 0.394)^T$.

The number of regressors equals three ($k = 3$). Firstly we consider a case of a small sample, when $n = 15$ observations take place. The regressors' values are the following:

$$X = \begin{pmatrix} 1 & 1 & 1 & 1 & 1 & 1 & 1 & 1 & 1 & 1 & 1 & 1 & 1 & 1 & 1 \\ 4 & 5 & 7 & 3 & 8 & 2 & 3 & 9 & 5 & 4 & 6 & 3 & 2 & 5 & 7 \\ 1.1 & 2.5 & 4.9 & 0.9 & 3.4 & 2.4 & 1.9 & 4.1 & 4.9 & 2.6 & 3.6 & 2.9 & 1.6 & 3.5 & 1.9 \end{pmatrix}^T$$

The two vectors $t$ and $I$ contain values of the durations of the observations and initial states of the environment:

$$t = (5 \ 8 \ 3 \ 6 \ 9 \ 6 \ 4 \ 6 \ 9 \ 8 \ 5 \ 7 \ 8 \ 10 \ 5)^T,$$
$$I = (1 \ 0 \ 2 \ 2 \ 1 \ 1 \ 0 \ 0 \ 1 \ 2 \ 1 \ 1 \ 0 \ 0 \ 2)^T.$$

Note that the total observation time equals to 99.

A simulation has been used for our purpose. The following parameters of the regression model are supposed: $\sigma = 1$,

$$\beta = \begin{pmatrix} 0 & 2 & 4 \\ 1 & 3 & 6 \\ 2 & 5 & 8 \end{pmatrix}$$

so $vec(\beta) = (0 \ 1 \ 2 \ 2 \ 3 \ 5 \ 4 \ 6 \ 8)^T$.

The first run of the simulation model gives the following values of the final states of the environment (vector $J$) and responses (vector $Y$):

$J = (0 \ 2 \ 0 \ 2 \ 2 \ 0 \ 1 \ 1 \ 1 \ 2 \ 2 \ 1 \ 0 \ 1 \ 0)^T,$
$Y = (67 \ 230 \ 231 \ 174 \ 296 \ 126 \ 43 \ 226 \ 427 \ 187 \ 284 \ 179 \ 146 \ 346 \ 262)^T.$

The data obtained are used for estimating the parameters $\beta$, those being supposed as unknown. We begin with the estimates for a simple linear regression (2) with the three regressors from $X$. Formula (3) gives $\widetilde{\beta} = (2.060 \ 3.454 \ 5.090)^T$. Weighted residual square sum $R = 9816$ against $R = 23200$ for one with respect to the sample mean.

Now we exam the suggested approach. The values of conditional average sojourn times $\bar{t}_i = (t_{i,1},...,t_{i,m})$ have been calculated by formulas (14) – (16). They are given for all observations in Table 1.

TABLE I. Value of average sojourn time $\bar{t}_i^T = (t_{i,1}, t_{i,2}, t_{i,3})^T, i = 1,...,15$

$$E(T) = \begin{pmatrix} i & 1 & 2 & 3 & 4 & 5 & 6 & 7 & 8 & 9 & 10 & 11 & 12 & 13 & 14 & 15 \\ j=1 & 1.8 & 3.2 & 1.4 & 1.5 & 1.9 & 2.1 & 1.7 & 2.6 & 1.7 & 2.2 & 0.7 & 1.1 & 4.5 & 4.0 & 2.3 \\ j=2 & 2.0 & 1.2 & 0.1 & 0.3 & 3.4 & 2.3 & 2.0 & 2.7 & 5.9 & 0.7 & 2.2 & 5.2 & 1.0 & 3.9 & 0.2 \\ j=3 & 1.2 & 3.6 & 1.5 & 4.2 & 3.7 & 1.6 & 0.3 & 0.7 & 1.4 & 5.1 & 2.1 & 0.7 & 2.5 & 2.1 & 2.5 \end{pmatrix}$$

Formula (3) gives the following estimates:

$$\widetilde{\beta} = \begin{pmatrix} -10.843 & -2.785 & 3.168 \\ 5.013 & -8.492 & 8.452 \\ 8.099 & 18.666 & 0.857 \end{pmatrix}.$$

They are very far from true $\beta$ but weighted residual square sum is sufficiently less: $R = 4674$. Therefore in the considered example the suggested method gives inaccessible estimates of true $\beta$ but predicts values of the responses well. What is the cause of such a fact? The given number of the observation ($n = 15$) is insufficient for the given number (9) of estimated parameters! After all, a random environment with exponential distributed sojourn times creates a big randomness. It can be seen from a value of estimated variance $\widetilde{\sigma}^2$, calculated by formula (8): one equals 934.8 although $\sigma^2 = 1$. This difference is explained by an essential randomness.

A possibility of using the approach described for a prediction can be seen from the following reasoning. For the considered regression model and given data on $t$, $I$, $J$, and $\bar{t}_i = (t_{i,1},...,t_{i,m})$ from Table 1, the expectation of the responses is the following:

$E(Y) = (44.7 \ 193.3 \ 131.1 \ 122.9 \ 295.1 \ 54.9 \ 9.8$
$65.1 \ 99.9 \ 247.8 \ 148.7 \ 32.8 \ 70.9 \ 126.7 \ 155.5)^T.$

Let us use this expectation as observed responses $Y$. The considered approach gives the same result: the residual square sum equals $6.207 \times 10^{-4}$ only! On other hand the estimates of $\beta$ are insufficient ones:

$vec(\widetilde{\beta}) = (0.024 \ 8.091 \times 10^{-3} \ -0.022 \ 0.015 \ -4.27 \times 10^{-4}$
$-3.658 \times 10^{-3} \ 3.992 \ 5.989 \ 8.023)^T.$

Note that the estimates $\widetilde{\beta}_3 = (3.992 \ 5.989 \ 8.023)^T$ are very close to the true parameter values for the third environment state: $\beta_3 = (4 \ 6 \ 8)^T$.

Formula (3) gives $\widetilde{\beta} = (7.503 \ 3.485 \ -2.149)^T$ and the weighted residual square sum $8.25 \times 10^3$ that is sufficiently worse.

Further we consider a case of a big sample. We organize the following simulation experiment. The last includes $q$ independent blocks. Each block has the above described structure: the same random environment, observation number $n = 15$, regressors' number $k = 3$. The matrix of regressors, the initial state $I$ of random environment $J$ and the observations' times $t$ are chosen at random according to the following distributions. The expectation of the matrix of regressors coincides with the above obtained matrix $X$, all elements of the second and third columns are independent and uniformly distributed on intervals (-2, 2) and (-1, 1) correspondingly. The expectation of observations' times coincides with previous values $t$, all times are independent and time of $i$-th observation $t_i$ has uniform distribution on (0, $2 t_i$). The initial states $I$ for various observations are independent and are chosen with respect to stationary distribution of the states for the random environment $J$. Therefore the total number of the observation for one experiment equals $qn = 15 q$.

The above described simulation procedure is realized for those conditions. Table 2 contains the obtained estimates of true parameters $vec(\beta) = (0\ 1\ 2\ 2\ 3\ 5\ 4\ 6\ 8)^T$ for increasing values of $q$.

TABLE II. RESULTS OF SIMULATION EXPERIMENT

| q | 500 | 1000 | 1500 | 2000 | 2500 | 3000 | 3500 | 4000 | 4500 |
|---|---|---|---|---|---|---|---|---|---|
| $\beta_{1,1}$ | 0.112 | 0.777 | 0.058 | -0.02 | 0.179 | 0.154 | 0.239 | 0.400 | 0.367 |
| $\beta_{2,1}$ | 0.711 | 0.770 | 0.881 | 0.898 | 0.992 | 0.868 | 0.851 | 0.888 | 0.948 |
| $\beta_{3,1}$ | 2.566 | 2.158 | 2.289 | 2.276 | 2.207 | 2.255 | 2.201 | 2.090 | 2.012 |
| $\beta_{1,2}$ | 2.746 | 1.572 | 2.170 | 2.306 | 2.105 | 2.092 | 2.026 | 2.047 | 2.061 |
| $\beta_{2,2}$ | 3.030 | 3.164 | 3.035 | 3.014 | 3.025 | 3.050 | 3.063 | 3.034 | 3.011 |
| $\beta_{3,2}$ | 4.707 | 4.932 | 4.935 | 4.913 | 4.948 | 4.928 | 4.934 | 4.960 | 4.978 |
| $\beta_{1,3}$ | 3.496 | 3.568 | 3.634 | 3.572 | 3.556 | 3.776 | 3.841 | 3.789 | 3.738 |
| $\beta_{2,3}$ | 6.152 | 6.079 | 6.058 | 6.061 | 6.113 | 6.105 | 6.099 | 6.071 | 6.079 |
| $\beta_{3,3}$ | 7.969 | 8.053 | 7.964 | 8.005 | 7.904 | 7.854 | 7.877 | 7.910 | 7.903 |

We see from Table 2 that a convergence to true values of parameters is very slow. One can be increased if an estimation procedure uses true values of variances for sojourn times. It will be a direction of our future investigations.